\documentstyle[twoside,fleqn,espcrc2]{article}
\input epsf
\newcommand{\pho}{\tilde{\gamma}}
\newcommand{\gl}{\tilde{g}}
\newcommand{\sneu}{\tilde{\nu}}

\newcommand{\sq}{\tilde{q}}

\newcommand{\gsi}{\,\raisebox{-0.13cm}{$\stackrel{\textstyle>}
{\textstyle\sim}$}\,}
\newcommand{\lsi}{\,\raisebox{-0.13cm}{$\stackrel{\textstyle<}
{\textstyle\sim}$}\,}
\newcommand{\be}{\begin{equation}} \newcommand{\ee}{\end {equation}}

\newcommand{\ttbs}{\char'134}
\newcommand{\AmS}{{\protect\the\textfont2
  A\kern-.1667em\lower.5ex\hbox{M}\kern-.125emS}}

\title{Status of Light Gaugino Scenarios}

\author{Glennys R. Farrar\address{ Department of Physics and Astronomy,
 Rutgers University, Piscataway, NJ 08855, USA}\ttbs{Research supported
 by NSF-PHY-94-2302.}}
       
\begin{document}

\begin{abstract}
I summarize recent devlopments in supersymmetry scenarios which leave
some or all gauginos light.  The emphasis is on experimental and
phenomenological progress in the past year.
\end{abstract}

\maketitle

\section{Introduction and Preliminaries}

Since SUSY95 there has been considerable interest in the possibility
of light gauginos.  It would be impossible to review here all
of the works on the subject, so I will limit myself to the following:
\begin{itemize}
\item  Attractive new models with a richer spectrum of gaugino masses.
\item  Higher order pQCD results including light gluinos.
\item  Constraints on the gluino mass and condensate from properties of the
$\eta'$. 
\item  New constraints from cosmology.
\item  A direct search for a decaying ``glueballino".
\item  A direct search for a decaying $R$-baryon.
\item  Limits from renormalization group running of $\alpha_s$,
extracted from the $e^+ e^-$ total hadronic cross section and $R_\tau$.
\item  Limit from a combined fit to properties of hadronic final states
in $Z$-decay. 
\item  Proposal for a model independent test of the all-gauginos-light
scenario at LEP2.
\item  Squark mass limits when gluinos are light.
\item  Some tantalizing observations which are readily explained with
light gluinos, but have eluded satisfactory explanation without them.
\end{itemize}

For many years, views about the possible superpartner spectrum were based
on models in which the fundamental interactions respect a GUT symmetry and
SUSY breaking is transmitted to the observable sector by gravity.  In such
models, one expects the tree-level masses of standard model gauginos to be
degenerate at short distance and diverge from one another at longer
distance because of renormalization group running, just as the gauge
couplings do.  Depending on the SUSY breaking mechanism, R-invariance
is strongly broken or not, leading to $m_{1/2} \approx M_{sq}$ or
$m_{1/2} \approx \frac{M_{sq}^2}{M_{Pl}}$.  In the latter case, the
low-scale gaugino masses are predominantly due to stop-top and
electroweak gauge-Higgs/ino loops\cite{betal}.  The resultant gaugino
masses are typically of order 1 GeV or less\cite{f:96,f:101}. For most
parameter choices the gluino has mass of order 100 MeV and the
lightest neutralino (a photino) is somewhat heavier due to significant
contributions of the electroweak loops\footnote{This observation is crucial
to the success of the light gaugino scenario.  Otherwise, the glueballino
lifetime would be so short as to have already been
excluded.}\cite{f:96,f:101}.  Charginos and heavier neutralinos get
their masses through  the Higgs phenomenon and mixing.  One chargino is always
lighter than the $W$ and three neutralinos lighter than the $Z$ in this type
of scenario, unless $\mu$ is very large.

In gauge-mediated models of SUSY breaking, gaugino masses arise from
1-loop interaction with messenger supermultiplets carrying the
gauge charge.  Superpartners of quarks and leptons get masses at two 
loops due to their coupling to gauginos.  If the messengers form a complete
GUT multiplet all gauginos are massive and their spectrum is like that in
SUGRA models.  However if the messengers are neutral under some
gauge group, those gauginos are massless at leading order.  Since quarks
and leptons have both $SU(2)$ and $U(1)$ charges, it is enough for the
messengers to have $SU(2)$ or $U(1)$ charges to make the squarks and
sleptons massive.  In this case, gluinos do not receive mass at 1-loop.
Wino and bino are massive or not, depending on the messenger
charges.  One can also obtain models in which all the standard model gauginos
are massless at one loop by introducing additional gauge interactions under
which the messengers and quarks and leptons are charged, but the messengers
are standard model singlets.  Mohapatra and Nandi (also
joined by Chacko and Dutta) and Raby have recently given models of these
types\cite{gmsb}.  One  interesting feature of the  
Raby model is that for part of parameter space the gravitino is heavier than
the gluino, as are the bino and wino, so the gluino is the LSP and is
stable.  Its mass can be adjusted over a large range, from $\approx 0$
to hundreds of GeV.

The new models were constructed to have nice properties in their own right,
such as solving the strong CP problem in the case of some of the
Mohapatra-Nandi models.  And as pointed out in \cite{f:103}, models
with sufficiently small dimension-3 SUSY breaking automaticalloy solve the
SUSYCP problem.  Although it appears possible to generate essentially arbitrary
combinations of gaugino masses, I will concentrate here on the following
different light-gaugino 
phenomenologies: 
\begin{itemize}
\item  All gauginos are nearly massless and get masses of order 1 GeV or
less from stop-top and electroweak loops.
\item  The gluino and the bino or wino is light.
\item  The gluino is the LSP.
\end{itemize}
The case of light wino and bino and heavy gluino is already excluded.  This
will be discussed below. 

Besides these model building accomplishments, the last two years have seen
advances in extending perturbative QCD in phenomenologically important
ways. The tour-de-force 1-loop calculation of 4-jet matrix elements in $e^+
e^-$ annihilation has been completed\cite{nlo} and applied to study of
event-shape and 4-jet angular distributions by Dixon and
Signer\cite{dixon_signer:R4,signer:moriond97}.
In addition, the 3-loop calculations of $R_{had}$ and $R_\tau$ and the
QCD beta-function have been extended to the case with light
gluinos\cite{clavho}.  These successes give hope that
the most daunting calculation of all, the 2-loop corrections to the 3-jet
matrix elements, will also prove feasible.  These results are
necessary for theoretical predictions to be sufficiently accurate to
discriminate the theory with and without light gluinos, as discussed
in sec. \ref{aleph}.

I now turn to experimental and phenomenological constraints on the
allowed mass range of gluinos.  The lightest hadron containing a
single gluino is expected to be the gluon-gluino bound state, usually
denoted $R^0$ (glueballino).   The relation between gluino current
mass and the mass of the $R^0$ can only be estimated.  A massless
gluino would imply a degenerate chiral supermultiplet consisting of
scalar, pseudoscalar and $R^0$, if mixing with $q \bar{q}$ states can
be ignored.  Quenched QCD predicts a $0^{++}$ glueball mass of 1.5-1.7
GeV and good candidates are the $f_0(1500)$ and
$f_0(1700)$\footnote{See \cite{f:109} for a recent discussion and
references.}.  Thus a massless gluino suggests an $R^0$ mass of about
$1 \frac{1}{2}$ GeV.  Its lifetime should be in the $10^{-5}-10^{-10}$
sec range, or longer if the up and down squarks are very
heavy\cite{f:101}.   

If the $R^0$ lifetime is less than about $10^{-10}$ sec, the original
missing energy\cite{f:24} and beam dump\cite{f:23} techniques are
applicable.  These methods have been used to establish $m_{\gl} \gsi 150 $
GeV\cite{pdg96}.  Ref. \cite{f:95} compiled experimental limits
relevant to gluinos for $R^0$ lifetimes longer than about $10^{-10}$ sec.
The most important constraint is from the CUSB
experiment\cite{cusb} which did not find $\Upsilon 
\rightarrow \gamma \eta_{\gl}$ at the pQCD-predicted level.  This
result rules out gluinos in the mass range $\sim 1\frac{1}{2} -
3\frac{1}{2}$ GeV and hence $R^0$'s between about $2\frac{1}{2} - 4$
GeV, for any lifetime\cite{f:95}.  As noted above, for $m(R^0)$ below
the CUSB-excluded range and photinos in the expected mass range of
$\sim \frac{1}{2} - 1$ GeV, the $R^0$ lifetime falls out of the range
of applicability of beam-dump and missing energy experiments and
previous limits were weak.  

If the gluino is above the CUSB limit of about $3 \frac{1}{2}$ GeV,
the $R^0$ mass should exceed the gluino mass by the constituent mass
of a gluon, $\frac{1}{2}-1$ GeV, in analogy with heavy-light quark mesons.
The spectator approximation is applicable and the $R^0$ lifetime is
well-approximated by the free-gluino lifetime.  For a light photino
this is: $\tau \approx 2~ 10^{-14} (\frac{M_{sq}}{100~{\rm GeV}})^4
(\frac{4~{\rm GeV}}{m_{\gl}})^5 $ sec.  Thus $m_{\gl} > 3 \frac{1}{2}$
GeV is only viable if up and down squark masses are at least $\sim
700$ GeV, or if $R^0$ decay to a lighter neutralino is kinematically
forbidden.  

In the case that the gluino and lightest hadron containing it are stable, the
most important consideration is whether any new hadrons bind to
nucleons to produce new stable nuclei which accumulate near Earth.  If so,
limits on exotic isotopes give stringent
limits\cite{f:51,plaga,nardi_roulet,nussinov:lg}. 
In order to produce an interesting dark matter density, a heavy gluino must
have a mass too large to be consistent with properties of our
galaxy\cite{nardi_roulet,nussinov:lg}. 

A very light gluino requires a corresponding pseudogoldstone boson.
The $\eta'$ suits this role well if $m_{\gl} = 80 - 140 $ MeV and $<\lambda
\lambda> =-(0.15 - 0.36)~ {\rm GeV}^3$\cite{f:108}.  This mass lies within 
the range estimated from the top-stop loop\cite{f:101}, and this
condensate is consistent with the naive expectation $
(<\lambda \lambda>)^{\frac{1}{3}}~ \approx \frac{9}{4} (<q
\bar{q}>)^{\frac{1}{3}}$\cite{f:95}. 

If the photino is the LSP and R-parity is exact, it can provide relic 
dark matter.  For a radiatively-generated photino mass, i.e., of order 1
GeV or less, obtaining the correct dark matter abundance requires $r \equiv
m(R^0)/m_{\pho}$ to be in the range 1.3 - 1.55\cite{f:100,f:113}.   In this
case, interconversion of photinos and $R^0$'s keeps them in equilibrium until
the appropriate epoch.  Photinos would ``overclose" the universe if
$r>1.8$ unless their mass is greater than about 10
GeV\cite{f:100,f:113}.  It is non-trivial that the predicted $R^0$
lifetime range $10^{-5} - 10^{-10}$ sec, is consistent both with the
experimental limits\cite{f:95} and the lifetime as estimated from
requiring the correct dark matter abundance\cite{f:113}.  

\section{New Experimental Constraints}

To summarize the foregoing, SUSY breaking scenarios in which gaugino masses
are mainly radiatively generated by known particles and their superpartners
are naturally consistent with the requirements of dark matter, the $\eta'$
mass, and direct experimental limits as of 1995.  The glueballino mass
should be in the range $\approx 1.4 - 2.2$ GeV on theoretical grounds;
experiment requires it to be less than about $2 \frac{1}{2}$ GeV.  Dark
matter is correct if $1.3 < r\equiv \frac{m(R^0)}{m_{\pho}} < 1.55$; $r>1.8$
is ruled out by cosmology unless $m_{\pho} \gsi 10$ GeV. We now turn to constraints
from new direct and indirect experimental searches. 

\subsection{Direct search via decays}
\label{direct}
\begin{figure}
\vspace{9pt}
%\framebox[55mm]{\rule[-21mm]{0mm}{43mm}}
\epsfxsize = 75mm\epsffile{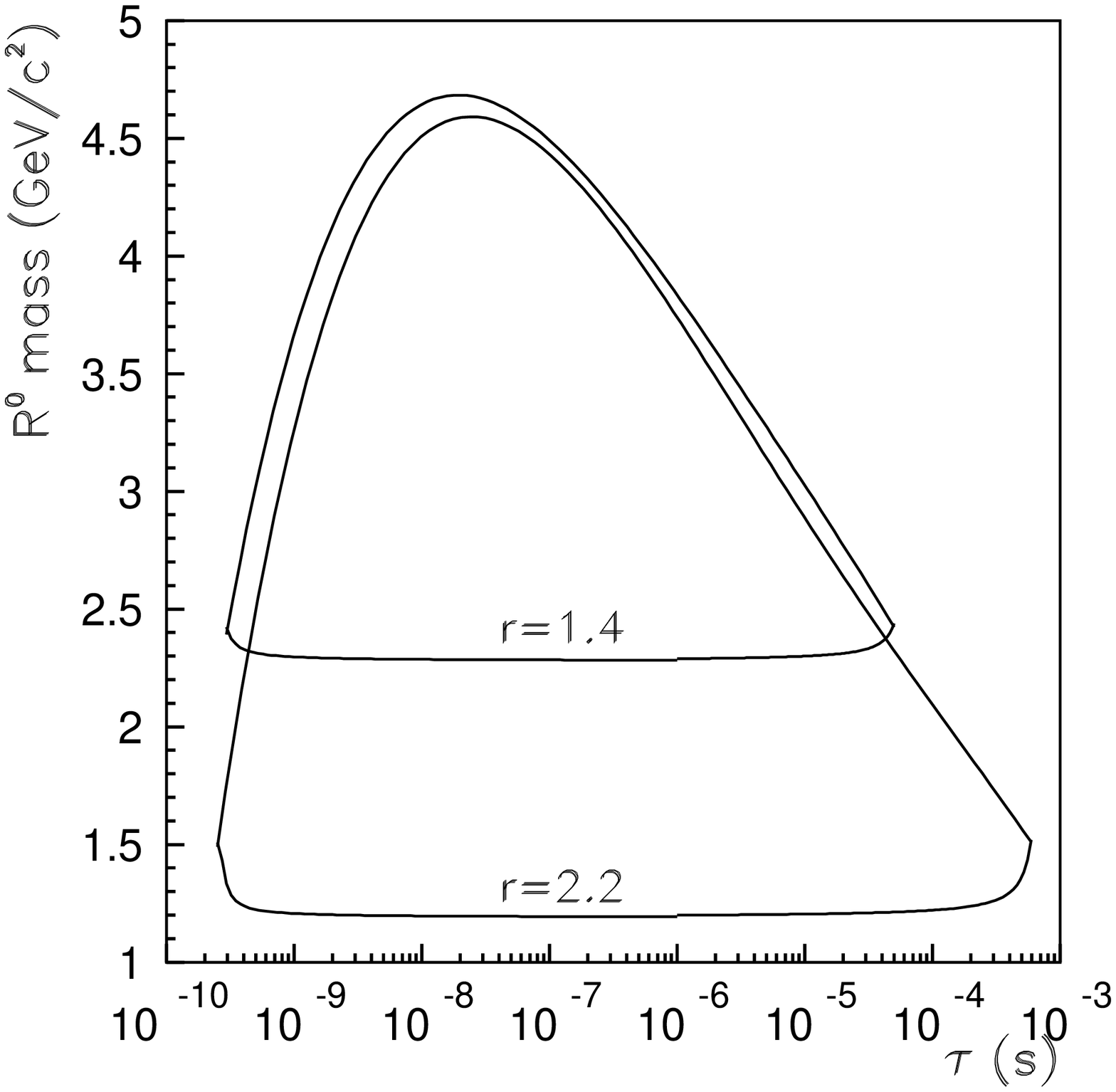}
\caption[]{KTeV limits\cite{ktev:lg}.} 
\label{ktev}
\end{figure}

The predominant decay mode of $R^0$'s in this scenario is to $\pi^+ \pi^-
\pho$\cite{f:104}.  Thus it was proposed that the current generation of
$K^0_L$ experiments search for evidence of $R^0$'s in their beam, whose decays
would result in $\pi^+ \pi^-$ pairs with high invariant mass and unbalanced
$p_t$\cite{f:104}.  KTeV has now completed such a study using a small fraction
of their total data\cite{ktev:lg}.  Their cut $m(\pi^+ \pi^-) > 648$ MeV
restricts them to the study of the kinematic region $m(R^0)(1-1/r) > 648$
MeV.  However subject to this constraint their limits are extremely good.
For the largest $r$ allowed by cosmology, 1.8, they are therefore sensitive
to $m(R^0)\gsi 1\frac{1}{2}$ GeV and considerably improve the previous
limits\cite{f:95}.  Fig. \ref{ktev} shows the mass-lifetime region excluded
by KTeV\cite{ktev:lg}, for two values of $r$.  Unfortunately the sensitivity
drops rapidly for lower $r$ and they are completely blind to the $R^0$ mass
region of primary interest, 1.4-2.2 GeV, for $r \le 1.4$.  The experimental
challenge will be to reduce the invariant mass cut.

%\begin{figure}[htb]
Another strategy to find a light gluino is to look in a charged hyperon beam
for the $R_p$ ($uud\gl$) decaying to the ground state R-baryon, the $S^0$
($uds\gl$)\cite{f:104}.  This weak decay, $R_p \rightarrow S^0 + \pi^+$, 
should have a lifetime $2 \cdot10^{-10}-2 \cdot 10^{-11}$ sec\cite{f:104}.
Assuming the gluino mass is $O(100~ {\rm MeV})$, the mass of the
$uud\gl$ state is calculated to be about 200 MeV higher than the
$S^0$\cite{f:51}.  The most interesting mass range for the $R_p$ is
$1.6 - 3.1$ GeV\cite{f:104}.  The lower end of this range is motivated
by the speculation that the $\Lambda(1405)$ is actually a crypto-exotic
flavor singlet $udsg$\cite{f:105}.  The upper end of this range is the 
mass above which the $S^0$ is likely to be strong interaction stable due to the
decay $S^0 \rightarrow R^0 \Lambda$.  The kinematics of the charged particles in this
decay is different from that of decays of known particles, so an upper
limit on the production rate of $R_p$'s can be obtained.  

Unfortunately, it is difficult to reliably estimate the production cross
section.  Perturbative QCD cannot be used, since the constitiuents of the
relevant state are not heavy and the transverse momentum is not large.
However one can use the measured $D$-meson differential cross section
as a benchmark for $R^0$ production, since their masses are comparable.
The larger color charge of the $R^0$'s constituents probably is not
relevant at low $p_t$ due to color screening.  On the other hand, no
known hadron provides a good analogy for $R_p$ production.  One would
expect the $R_p$ production cross section to be significantly lower 
than that of $\Omega^-$ or $\bar{\Xi}$, since the $R_p$ is heavier than these,
is pair produced with another particle whose mass is at least $1\frac{1}{2}$
GeV, and in addition requires binding 4 quanta rather than 3.  Hence upper
bounds on the production fraction of order $10^{-4}$ or better are probably
the minimum needed to hope to see a signal.  To rule it out, much better
limits would be needed.  

The $R_p$ search described above was performed by E761 at Fermilab and no
evidence for anomalous decays was found\cite{e761}.  The experiment's
best sensitivity is at $m(R_p) = 1.7$ GeV and $\tau(R_p) = 3 ~ 10^{-10}$
sec.  There, the sensitivity is about an order of magnitude lower than the
production level of $\bar{\Xi}$.  Given the suppression factors mentioned
above, this is a marginal level of sensitivity for exclusion.  However a
more serious problem is that the sensitivity drops rapidly as lifetime decreases
into the interesting range and as mass increases above about 2 GeV.  

%\begin{figure}[htb]
\begin{figure}
\vspace{9pt}
%\framebox[55mm]{\rule[-21mm]{0mm}{43mm}}
\epsfxsize = 75mm\epsffile{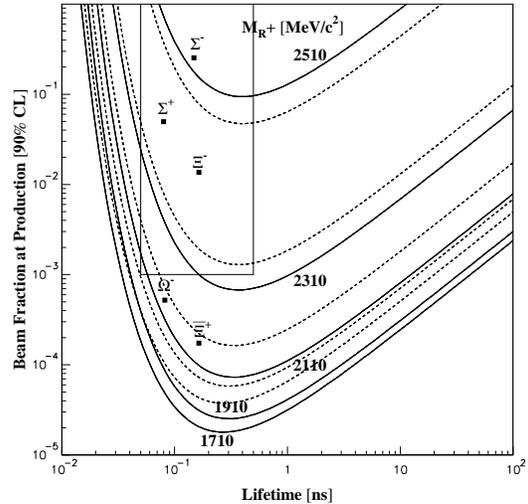}
\caption[]{E761 limits vs. $\tau(R_p)$\cite{e761} .} 
\label{e761}
\end{figure}
Thus the E761 search must be considered a first step which
demonstrates the feasibility of this technique.  I particularly want 
to stress that the $R_p \rightarrow S^0 \pi^+$ search is complementary to
the $R^0 \rightarrow \pi^+ \pi^- \pho$ search and both need to be pursued.
Since the former is a weak decay, the superparticle spectrum is essentially
irrelevant to the $R_p$ lifetime.  The lifetime is unaffected by whether
the gluino ($R^0$) is the LSP or whether the u- and d-squarks are light.
The search relies only on the mass of the gluino being low enough that $R_p$
production is at an experimentally accessible level.  On the other
hand, the $R^0$ search can be very sensitive to small production cross
sections because of the cleanliness and intensity of this generation of
kaon experiments.  However it depends in three crucial ways on the
superparticle spectrum:  i) If the u- and d-squarks are heavier than $\sim
130$ GeV, they must be heavier than $\sim 600$ GeV as discussed in sec. \ref{squarks}.
In this case the lifetime, $(10^{-7}-10^{-10})(\frac{M_{\sq}}{100~{\rm
GeV}})^4$ sec\cite{f:104}, may be inaccessibly long for
KTeV\footnote{The reaction $\pho \pi \leftrightarrow
R^0 \pi$ can still serve its crucial cosmological catalysis function due to the $R_\pi$ 
resonance\cite{f:116}.}. ii)  If the gravitino and lightest neutralino are
heavier than the $R^0$ it won't decay at all, assuming R-parity is good.
iii) If the $Q$ value of the decay is too small, because $R^0$ and $\pho$
are too close in mass, it will be difficult to discriminate signal from
background. 

Besides looking forward to improvements in the KTeV and E761 type
searches, the next couple of years may see other advances in direct
searches.  T. LeCompte and collaborators have been preparing a
parasitic search at Fermilab for states with anomalously long
lifetimes, which may be relevant for long-lived $R$-hadrons.  Also
Nussinov has made several suggestions on how to enhance the signal to
noise in searching for long-lived $R^0$'s\cite{nussinov:lg}.

\subsection{Running of $\alpha_s$}

Csikor and Fodor\cite{csikor_fodor} (CF) fit the $Q^2$ evolution
of $\alpha_s$ to the 3-loop beta function prediction, with and without
light gluinos.  The problem with previous attempts in this direction has
been the difficulty of determining $\alpha_s$ without reliance on models
of non-perturbative physics.  For instance a model or parameterization of
higher twist effects is required if deep inelastic scattering is used.  CF
employ $R_{had}$, the hadronic total cross section in $e^+ e^-$ annihilation.
This is the theoretically cleanest way to determine $\alpha_s$, on the 
assumption that we know the particle content of the theory.  Since it is a
total cross section it is insensitive to hadronization and does not require
resummation of large logarithms of the jet definition parameter $y_{cut}$.
Since it is known to 3-loops, it is insensitive to the arbitrary
renormalization scale, $\mu$. 

The drawback to using $R_{had}$ is that it is not very sensitive to $\alpha_s$, 
being proportional to $1 + \frac{\alpha_s}{\pi} + O(\alpha_s)^2$.  Thus the
statistical error in its determination is larger than in other methods. At
the $Z^0$ peak CF quote $\alpha_s = 0.123 \pm 0.006$ and at lower vales of
$\sqrt{s}$ the errors are very large.  Therefore CF also use $\alpha_s$ obtained
from the hadronic width of the $\tau$, which has a smaller statistical
error.  The perturbative contributions to $R_\tau$ are also known to
3-loops.  However the error on $\alpha_s(m_\tau)$ which should be associated
with modeling non-perturbative effects is controversial.  CF take $\alpha_s(m_\tau)
= 0.335 \pm 0.08$, but Shifman and collaborators argue that due to finite-size
singularities in the Operator Product Expansion (e.g., instantons) the actual
error may be twice as large as this\cite{shifman96}.  The need for a more
conservative error estimate is confirmed by Jan Fischer\footnote{Private
communication.  See also \cite{fischer:rev}.}. 

CF consider two mass regions for the gluino.  For $m_{\gl} < 1\frac{1}{2}$ GeV
they do not use $R_\tau$ in the fit and obtain a 70 \% cl exclusion
limit.  For $m_{\gl} = 3(5)$ GeV they do include $\alpha_s(m_\tau) =
0.335 \pm 0.08$ in the fit and obtain 95(90)\% cl exclusion limits.  However
had they used the larger error advocated in \cite{shifman96},
these limits would also be degraded to $ \le 70$ \% cl.  Thus unless the
theoretical uncertainty on $\alpha_s(m_\tau)$ can be 
reduced, or other sources of high-precicion, model-independent information
on $\alpha_s$ at low $Q^2$ can be found, the approach of using the
running of $\alpha_s$ to exclude light gluinos is inconclusive.

\subsection{Hadronic Event Shapes in $Z^0$ Decay}
\label{aleph}
The other new effort to exclude light gluinos is due to
ALEPH\cite{aleph:lg}.  Their strategy is to determine the effective number
of quark flavors, $n_f$, from a combined fit to several different 4-jet angular
distributions and $D_2$, the differential 2-jet rate.  At leading order,
adding gluinos to ordinary QCD increases $n_f$ determined via the running
of $\alpha_s$ or 4-jet angular correlations by 3 units.

The differential 2-jet rate is the number of events as a function
of $y_3$.  The variable $y_3$ is the value of $y_{cut}$ for a given event
at which it changes from being a 3-jet to a 2-jet event.  $D_2$ is
statistically powerful, since every hadronic $Z$ decay is used, but is quite
sensitive to the arbitrary renormalization scale, $\mu$, since it is only
known to 1 loop (NLO) accuracy.  Furthermore, the shape of the $D_2$
distribution is distorted in comparison to the fixed order PQCD
predictions due to logarithms of $y_3$ which are large when $y_3$ is
small.  These logs must be resummed, and the final prediction is dependent
on the procedure used to match the resummed 
and fixed order formulae.  The 4-jet angular distribution is
statistically weaker and is subject to large hadronization errors but
is insensitive to $\mu$\cite{dixon_signer:R4}. 
ALEPH's hope was that by performing a combined fit to both distributions,
their strengths might be complementary and their deficiencies less important.  

ALEPH reports $n_f = 4.24 \pm 0.29(stat) \pm 1.15(syst)$ and concludes that
they rule out a gluino with mass less than 6.3 GeV at 95 \% cl.  The precision
of the ALEPH result is governed by their systematic errors, specifically
the uncertainty in the theoretical prediction, since their statistical errors
are small.  Thus determining their systematic errors is the critical issue.
A detailed discussion can be found in refs. \cite{aleph:lg,f:118}.  Here
I give a brief synopsis of the main problems.

\begin{figure}
\vspace{9pt}
%\framebox[55mm]{\rule[-21mm]{0mm}{43mm}}
\epsfxsize = 75mm\epsffile{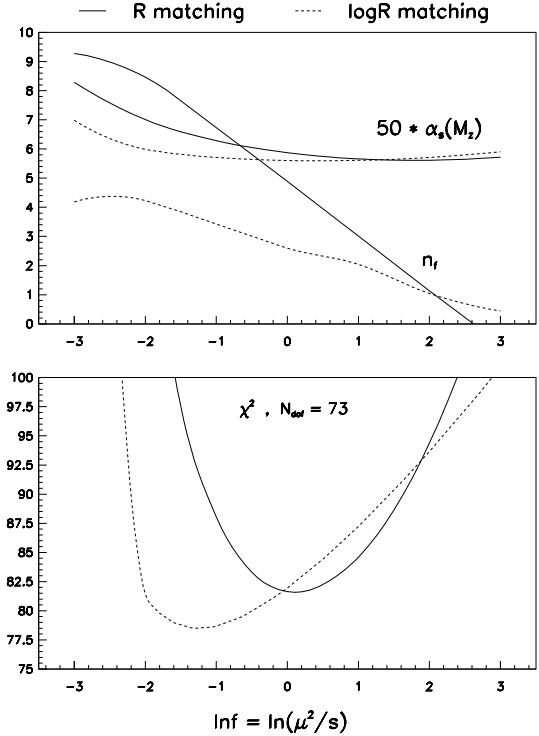}
\caption[]{ALEPH's $n_f$, $\alpha_s$, and $\chi^2$ vs $\mu$\cite{aleph:lg}.}
\label{aleph:mudep}
\end{figure}

The most important contribution to the systematic error is the uncertainty
in the predicted $D_2$ distribution due to truncation of the perturbation
series.  This is manifested by the sensitivity of the fit to renormalization
scale, $\mu$, and to the resummation scheme.  The conventional treatment 
of the uncertainty due to renormalization scale is to vary $\mu$ between
$\sqrt{s}/2$ and $2\sqrt{s}$, and treat the spread of results as a $\pm 1
\sigma$ error.  Such a procedure leads to a $\pm 2\frac{1}{2}$ unit
uncertainty on $n_f$ (see Fig. \ref{aleph:mudep}, from \cite{aleph:lg}).
With such a large uncertainty, no useful constraint on light gluinos can
be obtained.   

To reduce the uncertainty coming from scale sensitivity, ALEPH
assumed that there is a value of $\mu$ at which the resummed NLO prediction
for $D_2$ agrees with the all-orders prediction.  If this hypothesis is
correct, $\mu$ should be fixed to the value which gives the best fit, and the
associated uncertainty is essentially the statistical uncertainty in
finding $\mu$.  Thus the scale uncertainty in $n_f$ would be the range
in the $n_f$ obtained by varying $\mu$ to increase $\chi^2$ by 1 unit
above its minimum. 

Using the log R-matching scheme and varying $\mu$ to get the best fit, ALEPH
finds $n_f = 3.68$, with $\chi^2 = 78.5$ for 73 dof.  Increasing $\chi^2$
by one unit gives the range $n_f = 3.09 - 4.08$.  A similar exercise
for the R-matching scheme gives the best fit $n_f = 4.88$, with
$\chi^2 = 81.6$, and the range $n_f = 3.57 - 5.81$ as $\mu$'s
variation raises $\chi^2 $ to 83.  The other systematic errors they
estimate are $\pm 0.45$ from hadronization modeling and $\pm 0.27$ for
the detector simulation. Combining errors in quadrature and averaging
the results of the two matching schemes then gives their final result
quoted above. 

%\begin{figure}[htb]
\begin{figure}
\vspace{9pt}
%\framebox[55mm]{\rule[-21mm]{0mm}{43mm}}
\epsfxsize = 75mm\epsffile{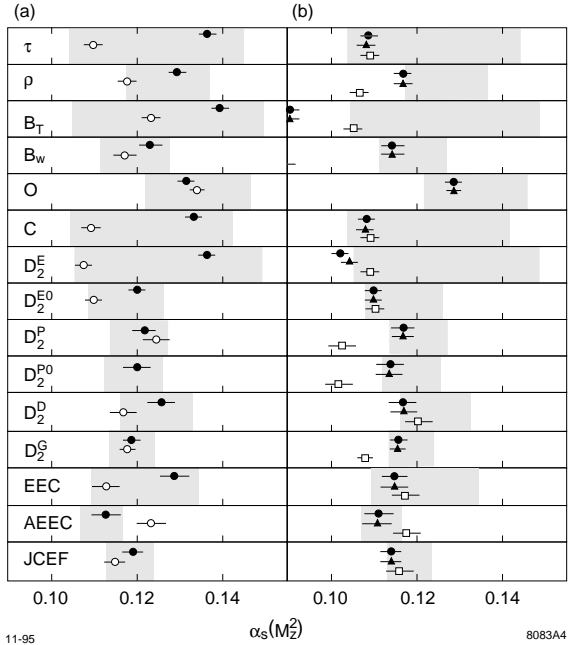}
\caption[]{(a) $\alpha_s$ from various event shape distributions using
$\mu = M_Z$ (solid circles) and experimental-optimization (open circles).
(b) Using other schemes\cite{burrows:warsaw}.} 
\label{burrows1}
\end{figure}

Unfortunately, the ``experimental optimization" procedure employed by ALEPH
to reduce their scale sensitivity is known to be invalid.
Burrows\cite{burrows:warsaw} examined the proposition that a judicious choice
of scale can improve the accuracy of the theoretical prediction for hadronic
event shape distributions.  He found that none of the scale fixing schemes,
including exerimental optimization, successfully improves perturbation
theory. His procedure was to extract $\alpha_s(m_Z)$ from 15 different
event-shape distributions, such as $D_2$, thrust, etc., using various
scale fixing schemes, such as experimental optimization, minimal
sensitivity, etc.  If a scheme provided a systemtic improvement, this
procedure would lead to a consistent set of $\alpha_s$ values, within
the errors from other sources. What Burrows found (see Fig. \ref{burrows1})
is that the dispersion in values is essentially the same in all schemes, and
comparable to the one found using the conventional
procedure\cite{burrows:warsaw}.  

%\begin{figure}[htb]

Thus ALEPH's method of fixing $\mu$ and estimating the theoretical
systematic error associated with the scale depedence is not valid.  In
the absence of a better way to estimate this uncertainty, the
conservative approach is to adopt the  traditional procedure used by
other experiments, which gives $\pm 2.5$.  At least this allows a
direct comparison between the sensitivities of previous analyses
employing just 4-jet angular distributions and the ALEPH procedure 
which also uses $D_2$.  This will be done below.

Another problem with the ALEPH analysis is the estimation of the
hadronization error.  That is in principle done by repeating the analysis
using several hadronization MC's which perform equally well for other purposes.
A fit using Herwig instead of Jetset gave $n_f = 6.21$ with $\chi^2 = 91.6$,
instead of $n_f=4.88$ and  $\chi^2 = 81.6$ (see Table 3 of \cite{aleph:lg}).
However $\mu$ was optimized only for the Jetset fit and that value was used
for all the different MC's, so there is no way of knowing what the Herwig
result would have been, or how small its $\chi^2$ would have been, had the
$\mu$ optimization been systematically applied.  Because the $\chi^2$ of
this Herwig fit is enough larger than for the Jetset best fit, the ALEPH
error-assignment procedure discards it entirely\cite{aleph:lg}.

It is also difficult to know what error to assign to the resummation
matching-scheme dependence of $D_2$.  ALEPH used the dispersion between log-R
and R matching. But since the log-R matching scheme gives $n_f < 5$ for any
value of $\mu$, we  can infer it is NOT the correct matching scheme.  

Even adopting the ALEPH hadronization error estimate and neglecting matching scheme
uncertainty altogether, one can see that employing $D_2$ with its strong
$\mu$ sensitivity leads to a worse determination of $n_f$ than using the
angular distributions alone.  If one uses the R-matching scheme result and
takes the $\pm 1 \sigma$ scale error to be given by the $1/2 < \mu/\sqrt{s}
< 2$ variation, the ALEPH result becomes $n_f = 4.88 \pm 0.29(stat) \pm 2.57(syst)$
which is considerably worse than the limits obtained by the other LEP groups
using just the 4-jet angular distributions\cite{lep4j}.

\section{Model Independent Proposal for a LEPII Search}

To summarize the previous section:  direct searches for decaying $R$-hadrons
have not yet explored the interesting regions of $m_{\pi\pi}$\cite{ktev:lg}
or $R_p$ lifetime\cite{e761}.  Indirect searches for light gluinos via $\alpha_s$
running\cite{csikor_fodor} and $Z^0$ event shapes\cite{aleph:lg} are stymied
by theoretical uncertainties.  In this section I describe a complementary
search technique\cite{f:116} which is theoretically very clean.  It is relevant
to the case that the gluino and either wino or bino is light;  it relies
on the high energy and integrated luminosity of LEP2.

When electroweak gaugino masses are negligible, the chargino and neutralino
masses depend only on\footnote{In this context, $\mu$ is the higgsino mass
parameter in SUSY models, not the renormalization scale!} $\mu$ and $tan
\beta$.  When $m_2$ vanishes, one chargino is lighter than the $W^\pm$.
Its mass decreases as $\mu$ or $tan \beta$ are increased, so large $\mu$
and $tan \beta$ are excluded by the $Z^0$ width\footnote{These statements
ignore radiative corrections, which is not strictly correct for extremely large
$\mu$\cite{f:96}.}.  However for small $\mu$, three neutralinos have masses
below the $Z^0$ if $m_1$ and $m_2$ are both small.  Considering the neutralino
contribution to the $Z^0$ width further restricts $\mu$ and $tan \beta$ but
does not exclude the scenario.

At energies above the $Z^0$, production of inos in $e^+ e^-$ collisions depends
on $\mu$ and $tan \beta$, and $m(\sneu_e)$ in the case of charginos.  Varying
$m(\sneu_e)$ over the allowed range, one can compute the minimum total cross
section for ino production, as a function of $E_{cm}$, $\mu$ and $tan
\beta$.  Even this lower limit is quite substantial:   2 pb at
173 GeV and 184 GeV, summing over chargino and neutralino production.

If the gluino is heavy, the charginos and heavier neutralinos decay mainly
into the lightest neutralino and products of $W^\pm$ or $Z^0$ decay. 
This possibility is already completely excluded by LEP\cite{fpt,opal:sighad}.
However if the gluino is light as well, inos can decay via a real
or virtual squark, to $q \bar{q} \gl$.  Indeed, this is the dominant
decay mode for squark masses up to $m_{sq} \sim 100$ GeV\cite{gf4j}.
Then the sensitivity of the usual ino searches is reduced by the factor
$(1-b)^2$, where $b$ is the suitablly averaged branching fraction of an ino
into $q \bar{q} \gl$.  However when both inos decay to $q \bar{q} \gl$, the
event contributes to the hadronic cross section.  The total hadronic cross
section is well-enough measured that $b=1$ is excluded for any value of $\mu$,
$tan \beta$, and $m(\sneu_e)$ by OPAL's recent analysis\cite{opal:sighad,f:116}.    
For the moment, the integrated luminosity is too small to extend the analysis
away from $b=1$ and $b=0$, but with the integrated luminosity planned at
183 GeV, LEP should be able to probe all values of $b$.  This method should
also permit the discovery or complete elimination of models such as Mohapatra
and Nandi's \cite{gmsb} in which only $m_2$ and $m_3$ are small, while $m_1$
and $\mu$ may be large. The possiblilty of putting interesting constraints
on other combinations is presently under investigation. 

A notable feature of this search is that it is theoretically very
clean\cite{f:116}. There is essentially no sensitivity to $\mu$.  The sensitivity
to parameters of the standard model such as $\alpha_s$ and $m_W$ is weak and
introduces an uncertainty which is small compared to present statistical
errors.  Most important, since it employs the total cross section the details
of hadronization are insignificant and there is no small parameter such
as $y_{cut}$ to introduce large logarithms which need resummation.

\section{Constraints on squarks if the gluino is light}
\label{squarks}

Lower bounds on squark masses are much weaker when the gluino is light
than when missing energy is a signature.  Early $Z^0$ width measurements
gave $m_{\sq} > 50-60$ GeV for degenerate
squarks\cite{Zsqlim}.  If only one flavor of squark is
light the limit becomes 30 GeV for a $L$-type squark and there is practically
no constraint for an $R$-type squark\cite{f:105}.  Data on the $Z^0$ has
improved enormously in the intervening period, inconsistencies with the SM
such as $R_b$ have disappeared, and higher energy data is now available.
Therefore the analysis of refs. \cite{Zsqlim} should 
be redone -- it would surely yield significantly better limits now.  

A stop more than a little lighter than the top would dominate top decay because
of $t \rightarrow \tilde{t} + \gl$.  This is probably excluded, but the possibility
that the stop decay products can ``fake" the final states observed by D0 and
CDF deserves investigation.

There have been a number of papers on the effect of associated squark-gluino
production on the dijet invariant mass distribution in $p \bar{p}$ collisions,
starting with Terekov and Clavelli\cite{clav:ltglsq}.  More recently, using
the full $106 pb^{-1}$ of Tevatron dijet data, Hewett et al\cite{hr:sqlim}
and Terekov\cite{t:sqlim} have extended the analysis also to dijet angular
distributions.  They are able to exclude the mass ranges $150 <
M(\tilde{u}) < 620$ GeV and $170 < M(\tilde{d}) < 620$
GeV\cite{t:sqlim} and $130 < M \lsi 600$ GeV\cite{hr:sqlim}.  At lower squark
mass, QCD background swamps the SUSY dijets; at larger mass the production
rate is too low.

Choudhury\cite{choudhury:mj} suggested that a monojet analysis might allow
Tevatron data to be used to exclude all squark masses below 240 GeV.  The
dijet-pair analysis proposed in \cite{f:105}, which should be useful for
lower squark masses and less model dependent than the monojet analysis, has
not yet been carried out. 

\section{Hints of a light gluino?}

There are several well-established phenomena, which have so far eluded
satisfactory explanation within the framework of standard model physics,
but which are readily explained with a light gluino.  

\subsection{The $\eta(1410)$}
The $\eta(1410)$ meson is a flavor singlet pseudoscalar.  All nearby pseudoscalar
nonets are filled, so that it must be a glueball or some other exotic
state\footnote{For a review and primary 
references see \cite{f:109}.}.  The relationship between its width and production
in $J/\psi$ radiative decay are apparently inconsistent 
with its being a $q \bar{q}$ state\cite{f:109}.  It cannot be identified
with a $K K^*$ "molecule" because no corresponding spin-1 state is
observed so $K K^*$ binding would have to occur only in the p-wave and not
in the s-wave which would be unprecedented.  Its coupling to glue-rich channels
is strong\cite{f:109} and it would naturally be interpreted as a glueball,
except that:    
\begin{itemize}
\item   For a 2-gluon state to have $J^{PC} = 0^{-+}$ requires one unit of
orbital angular momentum.  This suggests $m(0^{-+}) - m(0^{++}) \approx 
500-600$ MeV, the splitting between $^3P$ and $^1S$ mesons.
\item Lattice gauge calculations predict the mass of the lightest $0^{-+}$
glueball to be $2.2 \pm 0.3$ GeV; similar calculations for the $0^{++}$ and
$2^{++}$ sectors are within about 100 MeV or better of good candidate states. 
\item Models such as the bag and instanton gas calculations corroborate the
lattice result that $m(0^{-+}) >> m(0^{++})$. 
\item  There is some evidence for a suitable glueball candidate in the
mass range of the lattice prediction\footnote{W. Dunwoodie, private
communication.}. 
\end{itemize}
Overall, accounting for the $\eta(1410)$ and its strong affinity for
glue is difficult within QCD.  However the very-light gluino
scenario {\it predicts}\cite{f:95} the existance of a flavor singlet
pseudoscalar not present in conventional QCD, with mass about equal to the
unmixed scalar glueball, i.e., 1.4-1.8 GeV according to lattice gauge
theory.  The predicted properties of the $\eta_{\gl}$ fit the observed 
$\eta(1410)$ properties\cite{f:109}.

\subsection{Anomalies in jet production}  

Shortly after the announcement by CDF of an excess in very high $E_T$ 
jets\cite{cdf:hipjets}, several papers addressed the question of whether
light gluinos could account for the effect\cite{cdfjets}.
The excitement abated after CTEQ announced\cite{cteq4lq} that a slight
generalization of the functional form taken for the pdf's would allow the
gluon pdf to be increased enough to accomodate the new high $E_T$ jet data.
While the light gluino hypothesis improved the overall fit, it was not
essential.

However the most problematic anomaly in $p \bar{p}$ jet physics
is the strong violation of scaling observed by both CDF and D0 in the ratio
of $x_T$ distributions at $\sqrt{s} = 630$ GeV and $\sqrt{s} = 1800$ GeV.  Modifications
in pdf's change the inclusive $x_T$ distribution but do not significantly
affect the {\it ratio} of the scaled $x_T$ distributions.  As of the CTEQ
workshop in Nov. 1996, no explanation for the scaling violation had been
found within standard model physics.  Clavelli and Terekov\cite{clav:scaling}
point out that the observed breakdown of scaling may result from associated
production of a light gluino and squark, with $ M_{\sq} \approx 100-140$ GeV.

\subsection{Ultra High Energy Cosmic Rays}

Several cosmic rays of extremely high energy ($> 10^{20}$ eV) have been
detected.  Their shower properties are consistent with those expected
for a proton or nucleus primary, but not with a photon or neutrino
primary.  On the other hand, protons and nuclei of such high energy
interact strongly with the cosmic microwave background radiation via
the $\Delta$ resonance or nuclear breakup, leading to an upper bound on their
energies if they are to come from cosmological distances.  This is known
as the GZK bound\cite{gzk}.  Thus if the highest energy cosmic ray
$3 ~10^{20}$ eV were a proton, its source would have to be closer than about
50
Mpc\cite{elbert_sommers}.  The problem is that in order to produce such 
extremely high energy projectiles, the source is expected to be remarkable
in other observable ways.  In particular it should be an exceptionally strong
x-ray source\cite{hillas}.  There are no plausible candidates with appropriate
features in the relevant angular region, closer than 50 Mpc\cite{elbert_sommers}.
However the two highest energy cosmic rays point toward a good source
at about 240 Mpc, and an excellent one beyond 1000 Mpc\cite{elbert_sommers}.
It was suggested that these UHECR's might be the lightest $R$-baryon,
the $S^0$ ($uds\gl$) mentioned above\cite{f:104}.  Its 
interaction length in the CMBR is much longer than a nucleon's, and
can accomodate even a Gpc source\cite{f:114}.  Because its interactions with
atmospheric nuclei is similar to a nucleon's, its shower development
would be consistent with observations.  For further details,
references, and discussion of alternative "uhecrons" see \cite{f:114}.

\section{Summary}

The past two years have seen subtantial effort on many fronts to explore
the various light gaugino possibilities.  KTeV and E761 searches for evidence
of decaying $R$-hadrons have yielded null results, but they have not yet
investigated the most plausible regions of parameter space (see sec. \ref{direct}).

Indirect searches are still either theoretically or statistically 
inadequate.  The running of $\alpha_s$ as determined by $R_{had}$ in $e^+
e^- \rightarrow$ hadrons is theoretically clean but only gives a 
70 \% cl exclusion of light gluinos\cite{csikor_fodor}.  Analysis of $Z^0$
event shapes is statistically powerful but the theoretical predictions are
not known to high enough order, so that the renormalization scale and resummation
matching-scheme ambiguity is large.   In particular, ALEPH's claimed 
limit on light gluinos ($m_{\gl} > 6.3$ GeV at 95\% cl) must be set aside
because it relies on an ansatz for reducing the scale sensitivity which has
been shown to be invalid (see sec. \ref{aleph}).  Using a more realistic
estimate of the intrinsic theoretical uncertainty leads to the conclusion
that the ALEPH analysis is actually less sensitive than earlier
experiments employing only 4-jet angular distributions, for which systematic
uncertainties were also too large to make a definitive statement.  In
order to reduce the theoretical error in the ALEPH analysis to a useful level,
the two-loop correction to  three-jet matrix elements is needed.

It is still possible that all gauginos are massless, aside from radiative
corrections due to known particles and their superpartners.  This gives a
gluino mass is of order 100 MeV and is consistent with
properties of the $\eta'$.  Such a scenario provides an explanation
for dark matter, predicts the "extra" flavor singlet pseudoscalar 
($\eta(1410)$), and accounts for the apparent violation of the GZK bound
by ultra-high energy cosmic rays.  This scenario should be completely  
excluded, or suggestive evidence for it found, in the next year of LEP running,
using a combination of conventional signatures and limits on an excess in
the hadronic total cross section\cite{f:116,opal:sighad}.  With planned increases
in integrated luminosity, the case that only the wino and gluino are light
can be fully investigated in the same way.  The possibility that the wino
and/or bino are light, but the gluino is heavy, is already excluded by LEP. 

The most difficult case to study experimentally is when the gluino
is the only light gaugino.  With a significant improvement on the
determination of the $e^+ e^-$ total cross section at $Q^2 \ne m_Z^2$, 
the beta function can be determined with sufficient accuracy to unambiguously
infer or exclude some gluino mass ranges.  The requisite 3-loop calculations
have been done and the method does not require knowledge of non-perturbative
physics.  When the 3-jet differential cross section in $Z^0$ decay has
been calculated to 2-loop accuracy, it will probably be possible to use $Z^0$
event shapes to obtain a consitent set of determinations of $\alpha_s(m_Z)$
and $n_f$.  However further control of hadronization and resummation
uncertainties may also prove necessary.  A final strategy if the gluino
is not too heavy is to extend the search for the weak decay $R_p \rightarrow
S^0 + \pi^+$ in the fashion of E761\cite{f:104,e761} to shorter lifetimes
and much lower production levels.

%\bibliography{f,susy,qcd,cosmo}
%\bibliographystyle{unsrt}

\end{document}